\documentclass[twocolumn,aps,prb,showpacs,preprintnumbers,amsmath,amssymb]{revtex4}
\usepackage{graphicx}
\usepackage{bm}
\usepackage{gensymb}
\usepackage{color}
\begin{document}
\title{Size control of Charge-Orbital Order in  Half-Doped Manganite, La$_{0.5}$Ca$_{0.5}$MnO$_3$}
\author{Hena Das$^{1}$, G. Sangiovanni$^{2}$, A. Valli$^{2}$, K. Held$^{2}$ and T. Saha-Dasgupta$^{1,\ast}$}
\affiliation{$^1$ S.N. Bose National Centre for Basic Sciences,
Kolkata 700098, India} \affiliation{$^2$ Institute for Solid State Physics, Vienna University of Technology, 1040 Wien, Austria}
\date{\today}
\begin{abstract}
Motivated by recent experimental results, we study the effect of size reduction on half-doped manganite, 
La$_{0.5}$Ca$_{0.5}$MnO$_3$, using the combination of density functional theory (DFT) and dynamical mean field theory (DMFT).
We find that upon size reduction, the charge-ordered  antiferromagnetic phase, observed in bulk, to 
be destabilized, giving rise to the stability of a ferromagnetic metallic state. Our theoretical results, carried
out on defect-free nanocluster in isolation, establish the structural changes that follow upon size reduction to be 
responsible for this. Our study further points out the  effect of size reduction to be distinctively different from 
application of hydrostatic pressure. Interestingly, our DFT+DMFT study, additionally, reports the correlation-driven 
stability of charge-orbitally ordered state in bulk La$_{0.5}$Ca$_{0.5}$MnO$_3$, even in absence of long range
magnetic order.
\end{abstract}
\maketitle

Size  controls the physical properties of 
materials and can hence be employed to make materials functional. For strongly correlated materials, theoretical modelling of 
such phenomena is rare. Here, we pursue  such a study taking the case of half-doped manganites.
The charge and orbitally ordered state observed in half-doped manganites  
is one among the rich variety of fascinating phenomena exhibited by perovskite manganites R$_{1-x}$A$_x$MnO$_3$ (R= rare-earth element; 
A= alkali-earth element).\cite{1-2}  The charge-ordered (CO) state is 
associated with a real space ordering of Mn$^{3+}$/ Mn$^{4+}$ species in a 1:1 pattern. It is accompanied by orbital 
ordering (OO) and a structural change from orthorhombic to monoclinic symmetry which gives rise to an insulating 
ground state.\cite{3,4,5,7,8} 
The insulating CO state has been reported to be destabilized 
to a ferromagnetic (FM) metallic phase by various means that include a magnetic field,\cite{9}  doping, biaxial strain, pressure,\cite{10-11}  and
electric field.\cite{12-13}  Very recently, it has been shown in a few experimental studies\cite{14,15,new} that the destabilization of the CO state can be achieved even 
through size reduction. This interesting phenomenon adds another dimension, namely size, to the problem.
Size control is attractive from a technology point of view, which is achievable chemically in a low-cost way. 
The route through size control also opens up the possibility of exploring the 
tunability of the CO-OO state and the associated metal-insulator transition.

In this letter, we study the effect of size reduction on the CO-OO state of 
La$_{0.5}$Ca$_{0.5}$MnO$_3$ (LCMO) by using a combination of First-principles DFT and DMFT.
For the DFT calculations, we used a combination of two methods:
(a) plane wave-based pseudopotentials,\cite{vasp} and  
(b) muffin-tin orbital (MTO) based on linear muffin-tin orbital\cite{lmto} (LMTO) and N-th
order MTO (NMTO).\cite{nmto}  
For (a)  we used projected augmented wave pseudopotentials with an energy cutoff of 450 eV.
We used a spin polarized generalized gradient approximation (GGA).\cite{PBE} From a self-consistent DFT calculation, a low-energy Mn-e$_g$ only model Hamiltonian was 
constructed using NMTO-downfolding technique.\cite{nmto} The corresponding Hubbard Hamiltonian defined in
the downfolded NMTO basis was solved using DMFT, in the same spirit as previously carried out in 
in Ref. \onlinecite{Yamasaki06a} in the context of pure LaMnO$_3$. The low-energy model Hamiltonian consists of 
two $e_g$ orbitals per Mn ion with intra-orbital Coulomb interaction
$U=5\,$eV and Hund's exchange $J=0.75\,$eV which are coupled by ${\cal J}=1.35\,$eV to a (classical) spin representing the half-filled and immobile $t_{2g}$ electrons. The DMFT equations were solved by Hirsch-Fye quantum Monte Carlo simulations \cite{HF}, and, because of the CO ordering, it was necessary to explicitly consider a site-dependent double counting correction.\cite{Anisimov91}

Bulk LCMO shows a CO transition at T$_{co}$ = 155 K. Below 155 K, the crystal structure
is of monoclinic symmetry (P2$_{1}$/m) and an antiferromagnetic (AFM) order sets in.\cite{16}
The magnetic structure, the so called ``CE'' order,  consists of zig-zag FM chains that are coupled antiferromagnetically 
in the crystallographic $ac$ plane. The $ac$ planes are stacked antiferromagnetically along the crystallographic $b$-direction. A 
noteworthy feature of the crystal structure is that while the Jahn-Teller (JT) distortion is sizable for the bridge-site Mn 
atoms (Mn1) with two long bonds along the FM chain and four short bonds, the
corner-site Mn atoms (Mn2) on the zigzag chains have negligible distortion with nearly similar Mn-O bondlengths. Average Mn2-O 
distance is smaller than that of Mn1-O.\cite{3} Our DFT
calculations carried out on the experimentally measured structure,\cite{3} henceforth, referred as S$_{ex}$, showed 
the CE insulating phase
to be stable by 18 meV/f.u. over the FM metallic solution. The calculated electronic structure in terms of 
density of states and magnetic moments are found to be in good agreement with those reported previously in literature.\cite{17}

\begin{figure}
\includegraphics[width=8.5cm]{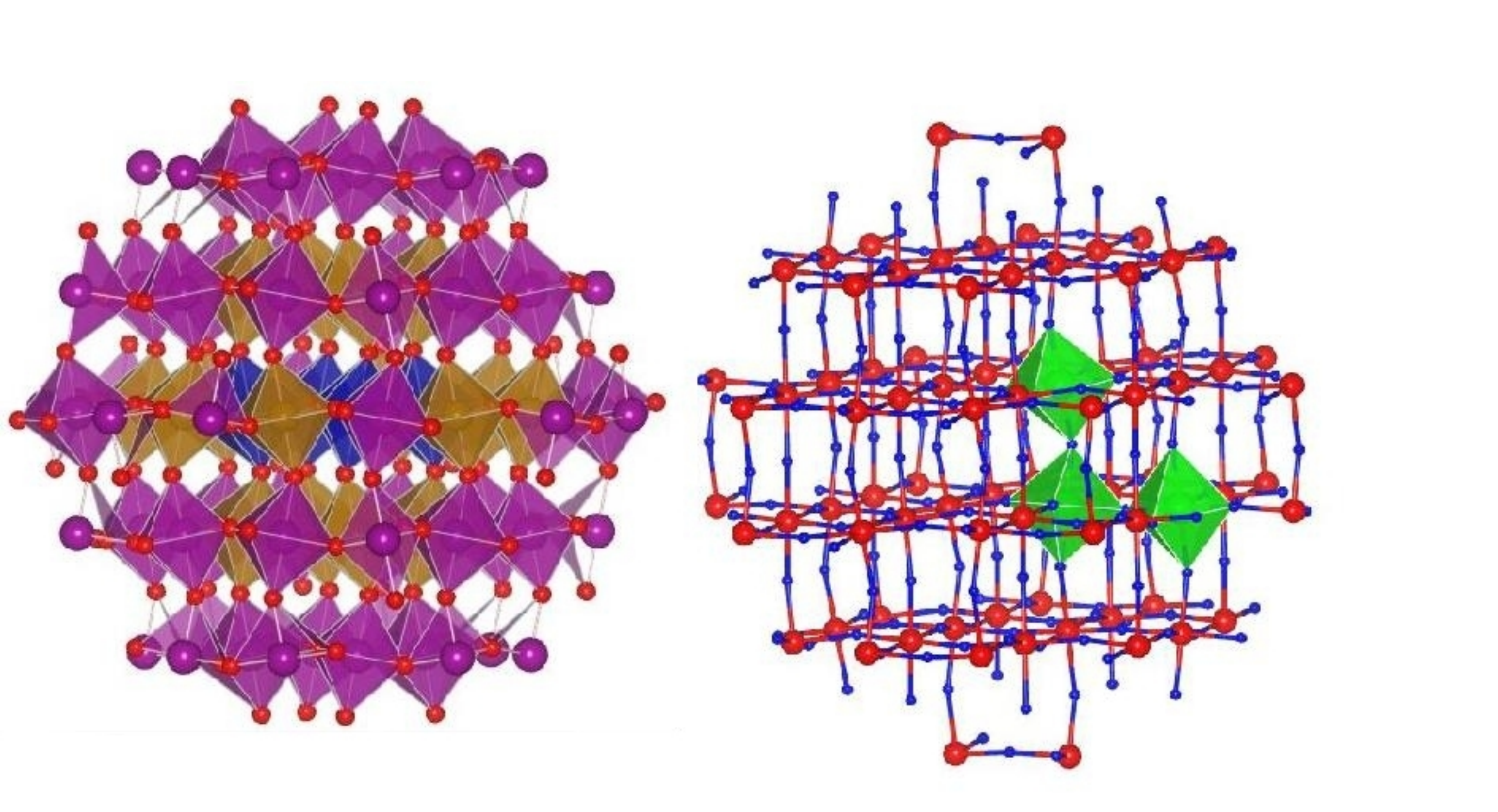}
\caption{(Color online) Left Panel: Constructed nanocluster of LCMO. In magenta (dark grey), brown (light grey) and blue (deep, dark grey) we show the MnO$_6$ octahedra belonging to the outer most surface layer, next to the surface and the core region.
Right Panel: The structural unit, highlighted in green (grey), chosen out of S$_{nano}$, used for building up of 
S$_{model}$. The big and small spheres in the unshaded region represent Mn and O atoms respectively.}
\end{figure}

In order to study the problem of nanoscale LCMO we first created a large supercell in the
monoclinic structure, from which a cluster of diameter 2 -3 nm having approximate spherical shape was cut out 
(cf Fig. 1, left panel), in which we carried out a full structural optimization. 
Following this procedure, the 2 nm cluster contains a total of 370 atoms and the 3 nm cluster contains a total of 700 
atoms, pushing it to a limit of our DFT structural optimization.  In the construction of the clusters, care has been taken 
to maintain the stoichiometry as closely as possible.  
For the cluster calculation, a simple cubic supercell was used with periodic boundary conditions, where two neighboring 
clusters were kept separated by 10 \AA, which essentially makes the interaction between cluster images negligible. The 
positions of the atoms were relaxed towards equilibrium, using the conjugate gradient technique
until the Hellmann-Feynman forces became less than 0.001 eV/\AA.

The considered DFT cluster sizes  are smaller than the experimental
realizations \cite{14} of sizes 15 nm. Hence only the inner region 
of the above constructed clusters of 2-3 nm size, is expected to mimic the 
prototypical behavior of the experimentally studied clusters.
In order to understand the consequences of the size-controlled
structural changes for such relatively larger clusters, 
we hence constructed a model bulk system, which we refer as S$_{model}$.  It is built 
out of the structural units belonging to the innermost core and the 
next to the core layer of the optimized LCMO in the nanoscale 
geometry, (referred as S$_{nano}$), as shown in the right panel of Fig. 1, and subsequently imposing
the symmetry considerations. 
The detailed procedure of construction of model system is explained in the
supplementary information (SI). Construction of S$_{model}$ leads to consideration
of the the local oxygen environments around Mn atoms as well as the tilt and 
rotation connecting two MnO$_6$ octahedra, same as that in core region of S$_{nano}$.
The lattice parameters and the Mn-O bond lengths of S$_{model}$ are compared 
to the bulk structure in Table I. The detailed structural information can be obtained in SI.
Note, for the bulk, we have considered the theoretically optimized structure, referred as S$_{bulk}$, 
in order to compare with parameters of S$_{model}$ in the same footing.
We find that the lattice parameters of S$_{model}$ show 
substantial reduction compared to those of the bulk system. The change in $a$ parameter appears to be the largest with a change of 
about 0.20 \AA, with moderate changes in $b$ and $c$ parameters, of 0.09 \AA's.  Qualitatively, this trend of reduction in 
lattice parameters and also the nature of reduction agrees very well with the crystal structure data extracted from X-ray diffraction 
of nanoclusters of  LCMO of 15 nm size [cf. Fig 4 in Ref.\onlinecite{14}(a)].  We note that the reduction in lattice parameters in the 
model structure gave rise to about 6$\%$ reduction in the volume compared to that of the bulk system; the first experiments\cite{14} 
report a 2$\%$ reduction. The 6$\%$ reduction was obtained for S$_{model}$ constructed out of S$_{nano}$ of 3 nm, while a similar procedure
for S$_{nano}$ of 2 nm, gave rise to larger volume reduction ($\approx$ 8 $\%$). This indicates the volume reduction to increase upon
decreasing cluster size, justifying the difference between the obtained volume reduction on 2-3 nm cluster and experimentally observed 
volume reduction on 15 nm cluster.
One of the important structural quantities is the orthorhombic 
strain: OS$_{\parallel} = 2\frac{(c-a)}{(c+a)}$ gives the strain in the $ac$ plane, while OS$_{\perp} = 2\frac{(a + c - \sqrt{2}b)}{(a+c+\sqrt{2}b)}$ is that along the $b$-axis. For S$_{bulk}$, the orthorhombic strain is highly anisotropic with a
negligible value of OS$_{\parallel}$ ($\approx$ 0.002) and a high value of OS$_{\perp}$ ($\approx$ 0.021). For S$_{model}$, we find 
instead the orthorhombic strains to be comparable (OS$_{\parallel}$ $\approx$ 0.02 and  OS$_{\perp}$ 
$\approx$ 0.01). This trend is also in very good agreement with experimental results.\cite{14} It confirms that our 
constructed model structure captures the essential structural changes in the nanoscale surprisingly well.
This proves that the role of surface beyond what is already taken into account in construction of model 
structure, is small.

\begin{table}
\begin{tabular}{|c|c|c|}
\hline
 & S$_{bulk}$ & S$_{model}$  \\ \hline
Lattice & $a$ = 5.47, $b$ = 7.58 & $a$ = 5.28, $b$ = 7.49 \\
param. & $c$ = 5.48 & $c$ = 5.39 \\ \hline
{\it Mn1-O:} & 2.18 \enskip 1.93 \enskip 1.94 \enskip 2.02  & 1.97 \enskip 1.92 \enskip 1.91 \enskip 1.93 \\
{\it Mn2-O:} & 1.92 \enskip 1.92 \enskip 1.94 \enskip 1.93 &  1.92 \enskip 1.88 \enskip 1.92 \enskip 1.91 \\ \hline
\end{tabular}
\caption{Lattice parameters, and Mn-O bondlengths (in \AA) of S$_{model}$ in comparison to S$_{bulk}$. 
The entries for Mn-O bondlength from left to right 
correspond to that along the FM chain, between the FM chains, along $b$-direction and the average. 
 Mn1 atoms are of nominal valence ${3+}$ and Mn2 of ${4+}$.\cite{foottable}}
\end{table}

Next, we calculated the electronic structure for S$_{model}$ and compared it with that of S$_{bulk}$.
For understanding the results, let us note that the difference between the average
Mn1-O and Mn2-O bondlengths is smaller in S$_{model}$ than in S$_{bulk}$. 
This leads to the expectation that the charge disproportionation (CD) between Mn1 and Mn2 sites to decrease in S$_{model}$. 
Furthermore, we note that for S$_{bulk}$, the difference between the longest
and the shortest Mn-O bondlengths is large for Mn1 and tiny for Mn2.
This gives rise to the crystal field splitting ($\Delta$)
between the two Mn-e$_g$ states, Mn-$3z^{2}-r^{2}$ and Mn-$x^{2}-y^{2}$, 
as large as 0.63 eV for Mn1 sites and less than 0.02 eV for the Mn2 sites.
In contrast for S$_{model}$, the bond length differences 
are much more similar for both types of Mn sites. This is reflected in similar 
$\Delta$'s for the nano model, i.e., 0.15 eV for Mn1 sites and 
0.10 eV for Mn2 sites. Together these two effects
weaken CO as well as OO in S$_{model}$.
This ordering is important to stabilize the  AFM structure found for the bulk. With  
charge and orbital ordering weakened, we find instead FM to be stable by 20 meV in S$_{model}$. This result 
is in accordance with the experimental observations.\cite{14,15} The microscopic 
origin of the size controlled transition from AFM to FM, therefore, can be traced back to the size-induced structural changes. 

\begin{figure}
\includegraphics[width=8cm]{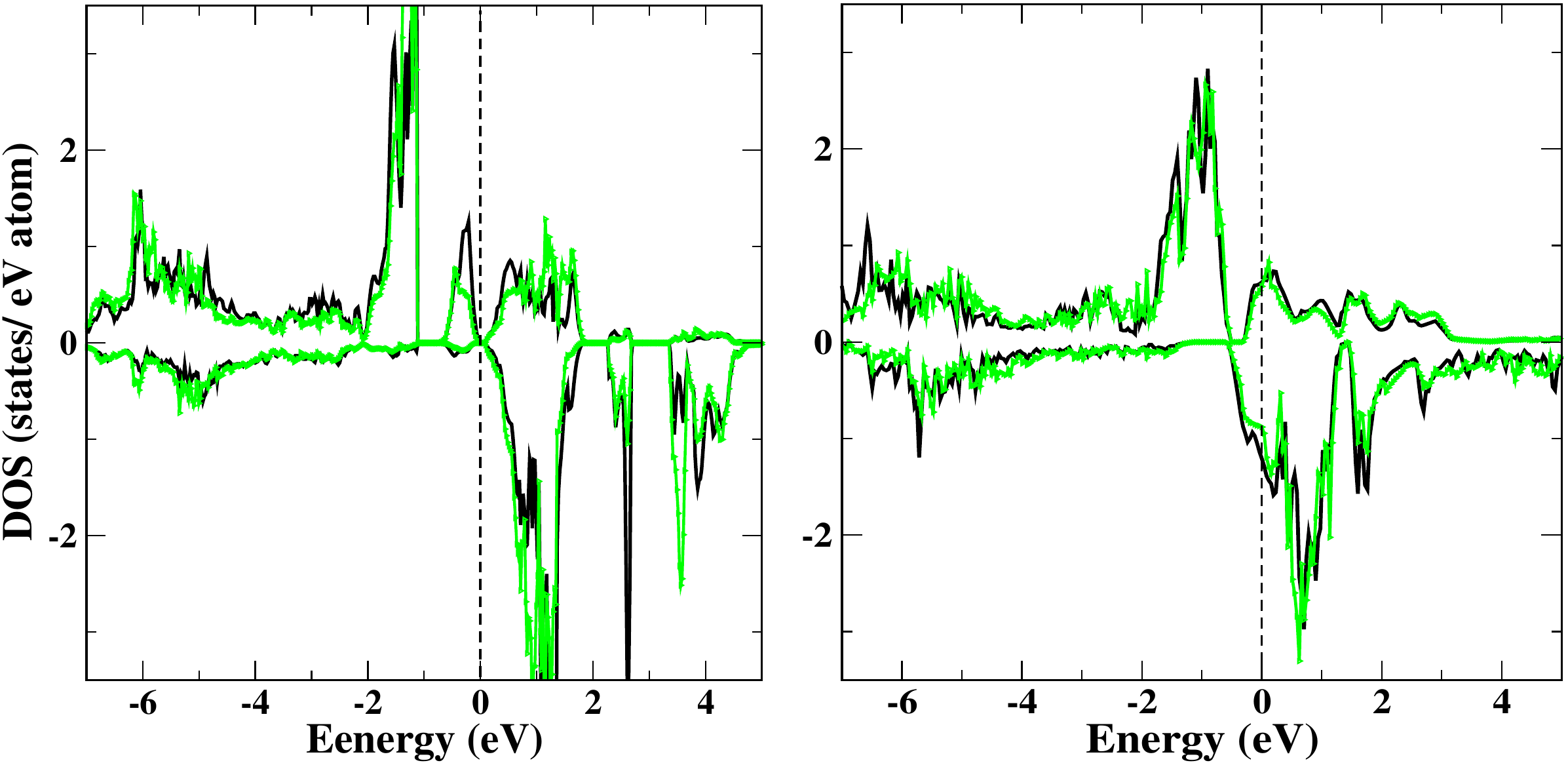}
\caption{(Color online)  DFT DOS, projected onto Mn1-$d$ (black solid line) and Mn2-$d$ (green/light grey line) 
states calculated for the CE phase of S$_{bulk}$ (left panel) and the FM phase of S$_{model}$ 
(right panel). The zero of  energy is set at E$_F$. \label{Fig:DOS}}
\end{figure}

Fig.\ \ref{Fig:DOS} shows the density of states (DOS)
of S$_{bulk}$ with AFM ordering of Mn spins, in comparison to that of S$_{model}$ with FM ordering.
Considering the DOS for S$_{bulk}$, the crystal field splitting between Mn-$3z^{2}-r^{2}$ and Mn-$x^{2}-y^{2}$
is clearly seen. In the majority spin channel, Mn-$3z^{2}-r^{2}$ states at Mn1 site 
are more occupied than the Mn-$3z^{2}-r^{2}$ states at Mn2 site, giving rise to CD between 
Mn1 and Mn2. We also find OO at Mn1 sites with a preferential occupation of Mn-$3z^{2}-r^{2}$ 
over Mn-$x^{2}-y^{2}$. The CO, although incomplete, together with the AFM spin ordering
gives rise to an insulating solution with a small but finite gap at E$_F$ already at the DFT level. 
Considering the DOS of S$_{model}$, 
we find that the splitting between Mn-$3z^{2}-r^{2}$ and Mn-$x^{2}-y^{2}$ is less pronounced and the
Mn1-$d$ and Mn2-$d$ states to be similar. 
The reduced $\Delta$ together with the increased bandwidth, compared to the bulk structure, drives 
S$_{model}$ to metallicity with a finite density of states at Fermi energy (E$_F$).
The increased bandwidth is caused by the  reduction in volume as well as by the FM ordering 
 which allows hopping processes within a double exchange model.\cite{de}

\begin{table}
\begin{tabular}{|c|c|c|}
\hline
 & Bulk & Nano-model  \\ \hline
Mn1(1) & 0.87 (0.50)  \enskip \enskip 0.01 (0.11) & 0.52 (0.31) \enskip \enskip 0.09 (0.20) \\
Mn1(2) & 0.85 (0.47)  \enskip \enskip 0.01 (0.12) & 0.72 (0.38) \enskip \enskip 0.04 (0.19) \\
Mn2 & 0.04 (0.15)  \enskip \enskip 0.09 (0.25) & 0.16(0.21) \enskip \enskip  0.16 (0.25) \\ \hline
\end{tabular}
\caption{Orbital occupancies for Mn-$3z^{2}-r^{2}$ (first entry) and $x^{2}-y^{2}$ (second entry) 
states calculated within DFT+DMFT for different inequivalent classes of Mn atoms in the unit cell 
for S$_{bulk}$ and S$_{model}$. 
In brackets we give the corresponding occupancies for the one particle, low-energy DFT Hamiltonian.}
\end{table}

\begin{figure}
\rotatebox{-90}{\includegraphics[width=6cm]{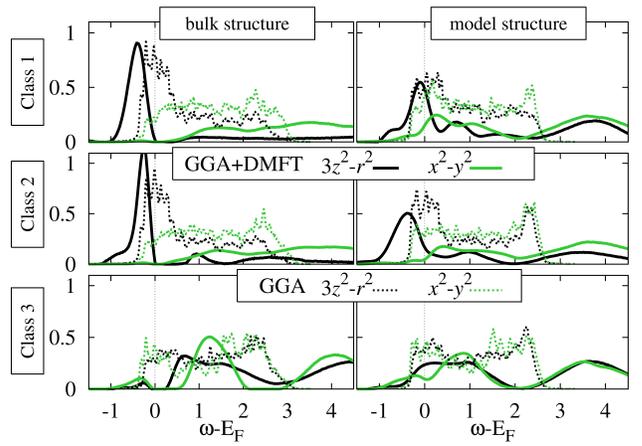}}
\caption{(Color online) DFT+DMFT spectral densities for e$_g$ states of three inequivalent classes of Mn (solid lines), corresponding to S$_{bulk}$ 
and S$_{model}$. The dashed lines represent the corresponding DFT DOS. The lines colored as black and green (light grey) correspond to $3z^{2}-r^{2}$ and $x^{2}-y^{2}$ states, respectively. \label{Fig:DMFT}}
\end{figure}

In order to take into account the influence of the missing electronic 
correlation in GGA, we did paramagnetic
 DMFT calculations for S$_{bulk}$ and S$_{model}$ structures, considering the low-energy Mn-$e_g$ only 
Hamiltonian derived out of DFT.
Table II lists the the orbital occupations of the three types
of inequivalent Mn sites. For S$_{bulk}$, already in the  DFT (in brackets) the two inequivalent 
Mn1 (``Mn$^{3+}$-like'') sites (Mn1(1) and Mn1(2)) are more occupied than the  Mn2 (``Mn$^{4+}$-like'') sites.\cite{footnote}
Besides the CD, there is, as mentioned above, a DFT orbital order. Electronic correlations 
enhance both kinds of ordering dramatically, making CO and OO nearly complete.
{\it This, establishes the correlation driven stability of CO and OO with
almost complete CD in a paramagnetic phase.}
The almost complete CD and  enhanced orbital polarization (OP) at the Mn1 sites
results in a gap at the chemical potential in the DFT+DMFT  spectral function for the bulk structure, as 
shown in the left panel of Fig.\ \ref{Fig:DMFT}, even without spin ordering. 
Note the opening of charge gap is stabilized by long-range Coulomb interactions which, within  
DMFT, reduce to their Hartree contribution. This effect is taken into account in our DFT+DMFT calculation on the GGA level.
Compared to the DFT spectra, spectral weight is transferred to high frequencies in the form of Hubbard bands, 
opening a gap at the chemical potential.
For the DFT calculation on the other hand, the charge 
disproportionation is incomplete, and the insulating solution is obtained only by assuming the AFM spin ordering.
Turning to S$_{model}$, the DFT occupancies show Mn$^{3+}$-like and Mn$^{4+}$-like sites to be similar 
with only a weak CD.  The inclusion of correlation 
effect through DMFT enhances CD to some extent following the trend seen for S$_{bulk}$. 
However, CD remains incomplete with an average occupation of  Mn$^{3+}$-like and  Mn$^{4+}$-like 
sites of 0.6-0.7 and 0.3 respectively, in comparison to 0.9 and 0.1 respectively, obtained for S$_{bulk}$.  
{\it This conclusively establishes that size reduction leads to weakening of 
charge disproportionation.} This in turn leads to 
metallic DFT+DMFT solution for S$_{model}$ with finite weight at the chemical potential, as shown in right panel of Fig.3.
Note, although S$_{nano}$ does not maintain the strict stoichiometry, the constructed S$_{model}$
is strictly stoichiometric, pointing the fact that destabilization of CO is driven by the structural changes due to size 
confinement, rather than due to deviation from half-doping.

As one of the major structural changes upon size reduction is the volume compression, it is worthwhile to
compare the structural and electronic changes induced by size reduction to those occurring under hydrostatic pressure.
To this end, we carried out calculations of LCMO, with uniformly reduced lattice
parameters with  6$\%$ reduced volume, the atomic positions being   
optimized in DFT, referred as structure S$_{press}$. The details of the optimized structure is given in SI. 
Following the selfconsistent DFT calculations on S$_{press}$, 
the Mn-e$_g$ only low-energy Hamiltonian was
constructed and the corresponding Hubbard Hamiltonian was solved using DMFT. Compared to S$_{model}$, first of all, 
we find that at the DFT level, CD and OP
is much weaker, even though the volume is the same. With this 
 less polarized starting point, all Mn sites are filled with $\approx 0.5$ electrons. In this situation,
electronic correlations are less relevant. The DMFT orbital and site occupations remain very similar to the DFT values with 
$0.4$ - $0.6$ electrons/site, and the system is far away from a metal-insulator transition (MIT).
This leads us to conclude that the nanoscopic system is much closer to a MIT than bulk La$_{0.5}$Ca$_{0.5}$MnO$_3$ 
under hydrostatic pressure.\cite{note} The size reduction and application of hydrostatic pressure, 
therefore, should be considered as two very different routes.

In conclusion, using DFT calculations combined with DMFT, we have studied the effect of size reduction on 
charge-orbital order in  half-doped LCMO manganites. Our study indicates that the size reduction leads to substantial 
reduction in volume as well as a change in the nature of the orthorhombic strain. The structural changes under size 
reduction lead to a weakening of both charge and orbital ordering, making the ferromagnetic metallic state 
energetically favorable compared to the ``CE'' type antiferromagnetic
insulating state, which is the ground state of the bulk structure. While such effect has been
observed, the experimental situation is faced with difficulties, like possible presence of impure
phases, the grain boundaries, non-stoichiometry. Our theoretical calculations were carried out considering
nanocluster in isolation, and therefore, devoid of such complications. Through construction of model structure,
the issue of non-stoichiometry was avoided.
Furthermore, the effect of size reduction turned out to be
very different
from that of pure hydrostatic pressure. We predict the nanoscopic system to be close to the MIT in comparison 
to the system under hydrostatic pressure with the same amount of volume reduction. Increasing the size of the nanocluster, one
would expect to drive the system closer and closer to MIT. It would therefore be possible to tune the LCMO system to the verge 
of MIT, and thereby, achieve a large magnetoresistive response under small magnetic fields. Finally, while
we carried out our investigation on LCMO, the destabilization has been predicted for Pr$_{0.5}$Ca$_{0.5}$MnO$_3$ too,\cite{pcmo} hinting to oberseved effect to be a more general one. This will be taken up in a later study.

We thank A.K. Raychaudhuri and T. Sarkar to bring this problem into our notice, the EU-Indian network MONANI and  SFB ViCom F41, GK W004 (AV) and  n.M1136 (GS) of the FWF for financial support. Calculations have been done on the Vienna Scientific Cluster.

\vskip .2in

$\ast$ tanusri@bose.res.in


\begin{thebibliography}{99}
\bibitem{foottable} There are three inequivalent classes of Mn atoms in the unit cell, the first two are of 
nominal valence ${3+}$ and the last one is of ${4+}$. The structural parameters corresponding to only the first Mn$^{3+}$ is shown. Also, for Mn$^{4+}$-type three bondlengths are shown, taking the average of the pairs in the
mentioned directions. For details see Table 2 of SI.
\bibitem{note} Although compression of volume and Mn-O bonds happen both 
in nanoscale and on application of pressure, the larger OS$_{\parallel}$, 
in case of nanoscale compared to that under pressure makes the
Mn-e$_g$ bandwidth smaller for the former, through reduction
in Mn-O-Mn bond angle.
\end{thebibliography}
\end{document}